# Engineering spin defects in hexagonal boron nitride

Mehran Kianinia,[1,*] Simon White,[1] Johannes E. Fröch,[1] Carlo Bradac,[1] and Igor Aharonovich.[1,2,*]

[1] School of Mathematical and Physical Sciences, University of Technology Sydney, Ultimo, New South Wales, 2007, Australia
[2] ARC Centre of Excellence for Transformative Meta-Optical Systems, Faculty of Science, University of Technology Sydney, Ultimo, New South Wales, 2007, Australia

[*]mehran.kianinia@uts.edu.au, [*]igor.aharonovich@uts.edu.au

## Abstract

Two-dimensional hexagonal boron nitride offers intriguing opportunities for advanced studies of light-matter interaction at the nanoscale, specifically for realizations in quantum nanophotonics. Here, we demonstrate the engineering of optically-addressable spin defects based on the negatively-charged boron vacancy ($V_B^-$) center. We show that these centers can be created in exfoliated hexagonal boron nitride using a variety of focused ion beams (nitrogen, xenon and argon), with nanoscale precision. Using a combination of laser and resonant microwave excitation, we carry out optically detected magnetic resonance spectroscopy measurements, which reveal a zero-field ground state splitting for the defect of ~3.46 GHz. We also perform photoluminescence excitation spectroscopy and temperature dependent photoluminescence measurements to elucidate the photophysical properties of the $V_B^-$ center. Our results are important for advanced quantum and nanophotonics realizations involving manipulation and readout of spin defects in hexagonal boron nitride.

## Main

Solid-state defects with optically addressable spin states are highly sought-after for the realization of quantum photonic devices and scalable quantum information architectures.[1-2] Significant effort has been devoted to engineer, characterize and control spin defects in diamond (e.g. nitrogen vacancy and silicon vacancy centers),[3-5] silicon carbide (e.g. di-vacancy centers)[6-9] and rare earth materials (e.g. Yb ions in yttrium orthovanadate hosts).[10-11] The undisputed success of these systems—marked by the ability to initialize, manipulate and optically read out individual spins with long coherence times—is however accompanied by ongoing challenges.[1-2] These include the ability to engineer quantum emitters which, besides exhibiting spin-dependent photon emission, also display high photon extraction rates and ease of integrability with other (hybrid) nanoscale systems.[12-14]

In this context, two-dimensional (2D) van der Waals materials, and specifically hexagonal boron nitride (hBN), have recently emerged as promising platforms for integrated nanophotonics.[15-16] Hexagonal boron nitride has been employed, for instance, in the realization of photonic crystal cavities[15, 17] and optomechanical resonators,[18] as well as in the study of fundamental phenomena involving the subwavelength propagation of phonon-polaritons.[19] The material is also host to atom-like quantum emitters.[20-23] The origin of these quantum emitters is still unknown.[24-27] Furthermore, their on-demand fabrication (both in terms of spatial accuracy and consistency of their spectral features) in exfoliated hBN flakes is still an outstanding goal.[23, 28] Yet, they are widely utilized due to desirable properties such as high brightness and stability,[29] large stark-shift tuning,[30-31] photo-physics compatible to super resolution imaging,[32] and addressable spin-dependent optical emission[33-36]—which is the focus of this work. The ability to fabricate and control stable emitters with optically-addressable spin states is, in fact, a key factor in aspiring to use hBN in spin-based sensing applications and quantum information technologies.

Here we demonstrate the controlled engineering of boron vacancy ($V_B^-$) defects that exhibit optically detected magnetic resonance (ODMR), at room temperature. We build upon the recent observation of ODMR in neutron irradiated bulk hBN crystals[33] and demonstrate a versatile method for creating the $V_B^-$ centers responsible for the ODMR signal. The method relies on focused ion beam (FIB) implantation and allows for the patterning of arrays of spin defects with nanoscale precision. Our results mark an important step towards the controlled generation of optically-active hBN color centers with spin-dependent photon-emission.

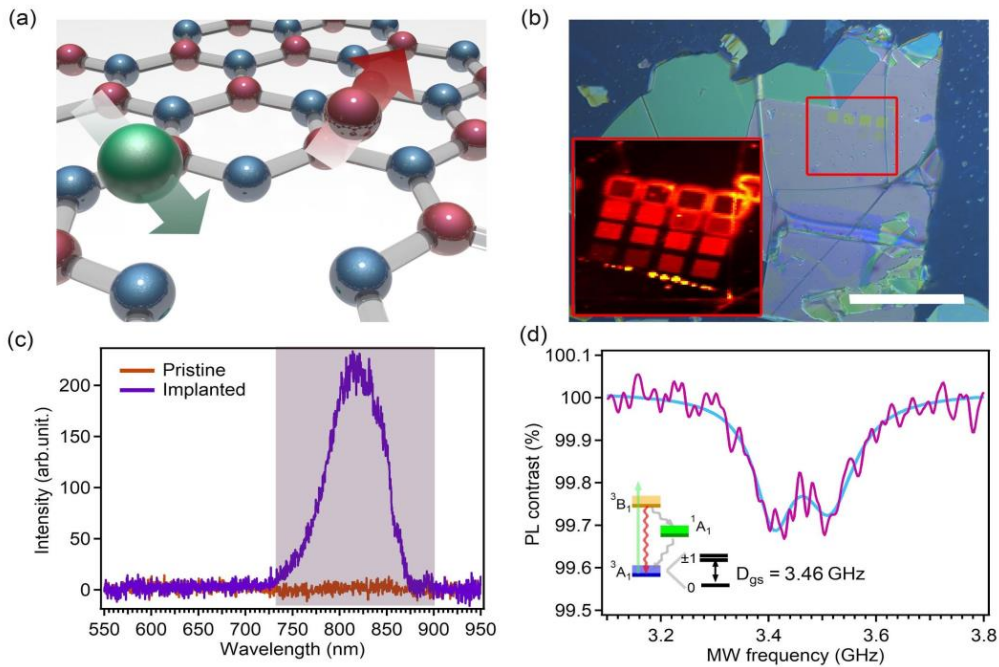

**Figure 1.** Creation of boron vacancy ($V_B^-$) defects in hBN. **a)** Schematic of ion implantation process. Xenon ions (green), knock out boron atoms (red) from the hBN lattice leaving behind ensembles of $V_B^-$ defects. **b)** Optical image of an exfoliated hBN flake after patterning. The highlighted area shows an array of patterned squares, with individual areas of 16 μm$^2$, with increasing Xe ion fluence. The inset shows a photoluminescence (PL) map of the patterned area. The scale bar is 20 μm. **c)** PL spectrum of the implanted area: the $V_B^-$ emission is centered at ~820 nm. **d)** ODMR measurement from the implanted area with no external magnetic field. The experimental data (purple) is fitted with two Lorentzian (light blue). Inset: level structure of $V_B^-$ with the measured zero-field splitting of ~3.46 GHz.

We created boron-vacancy ($V_B^-$) defects using ion implantation process in a commercially available focused ion beam (FIB) system. Hexagonal boron nitride crystals were tape-exfoliated and transferred onto a silicon substrate, which had a 300-μm-thick thermal oxide layer. The characterized hBN flakes were typically 100-200 nm thick. The sample was then moved into a FIB microscope and the hBN flakes were implanted with a $Xe^+$ beam accelerated to 30 keV. The process is schematically shown in Figure 1a. The high energy Xe ions break the B–N bonds in the hBN lattice and leave missing atoms (vacancies) behind. Figure 1b is an optical image of the hBN flake after the patterning with Xe ions. The implanted regions are clearly visible both in the optical and photoluminescence images (Figure 1b and inset). We patterned a matrix of 4×4 squares, each square being 16 μm² in size, with an increasing fluence from $1\times10^{14}$ to $1\times10^{17}$ ions/cm². The patterned areas are indicated with the red box in Figure 1b.

To characterize the patterned spots, we carried out a confocal scan of the area (Figure 1b, inset) using a 532-nm excitation laser. The implanted areas exhibit strong photoluminescence (PL) emission centered, spectrally, at ~820 nm. Conversely, the pristine areas do not display any PL signal (Figure 1c). The observed emission is characteristic of $V_B^-$ centers and confirms the selective creation of $V_B^-$ defects in hBN by the Xe ion implantation process. The intensity of the PL signal increases as the Xe fluence increases (see below, Figures 2a and 2b). We note that towards higher fluences, the Xe ions start causing sputtering and amorphization of the hBN material. This is visible in Figure 1b where a different color contrast in the optical (and confocal) image is observable as the material becomes thinner and damaged in the middle of the implanted area (see, e.g., the top four squares of the 4×4 matrix).

To verify that the created defects are $V_B^-$ centers, we performed optically detected magnetic resonance (ODMR) measurement, at room temperature. A thin copper wire (diameter 20-μm) was suspended in close proximity (~μm away) to the implanted area and used as an antenna to deliver a microwave field while the sample was excited with a 532-nm laser; no external magnetic field was applied. We monitored the PL emission (spectral range 720–900 nm) as we swept the frequency of the microwave field from 3.1 to 3.8 GHz. The results are shown in Figure 1d. The fluorescence signal drops when the microwave field oscillates at $\nu_1$ ~ 3.41 GHz and $\nu_2$ ~ 3.51 GHz. This is consistent with the spin-optical dynamics reported for $V_B^-$ defects in hBN.[33, 37] The $V_B^-$ defect has a triplet ($S = 1$) ground state with a zero-field splitting described by the parameters $D_{gs}$ and $E_{gs}$. In our experiment we find $D_{gs} \cong 3.46$ GHz and $E_{gs} \cong 50$ MHz, in Planck's constant units $h$. The triplet energy sublevels have a completely-lifted threefold degeneracy even in the absence of an external magnetic field ***B***, due to the small off-axial component $E_{gs}$. The resonant frequencies $\nu_1$ and $\nu_2$ in the ODMR spectrum are given by $\nu_{1,2} = D_{gs}/h \pm (1/h)\sqrt{E_{gs}^2 + (g\mu_B B)^2}$ where $g$ is the Landé factor and $\mu_B$ is the Bohr magneton. Our measurements (Figure 1d) were carried out in the absence of external magnetic fields, thus $\nu_{1,2} = (D_{gs} \pm E_{gs})/h$. They show the expected behavior reported for the $V_B^-$ center where optical excitation involving the excited and metastable states preferentially populates the ground spin sublevel $m_s = 0$, which, upon excitation, scatters more photons than the $m_s = \pm 1$ ground sublevels populated by the resonant microwave field.

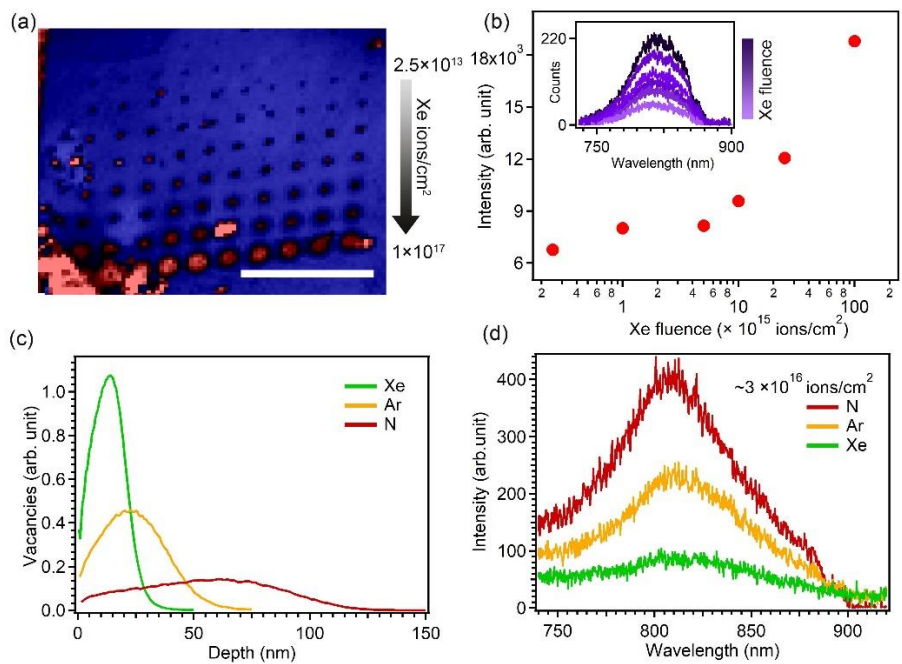

**Figure 2.** Engineering of $V_B^-$ defects in hBN. **a)** Confocal scan of the $V_B^-$ in hBN created with an increasing fluence of xenon ions (scale bar 20 µm). **b)** PL intensity as a function of the Xe fluence. Inset: corresponding PL spectra at each implantation fluence. **c)** SRIM simulation for implantation with Xe, Ar or N ions. **d)** PL emission at the same fluence of ~$3\times10^{16}$ ions/cm² for each of the implanted species that results in the creation of $V_B^-$ in hBN.

Next, we studied the effect of the implantation fluence and species on the formation probability and brightness of the created $V_B^-$ defects in hBN. To this end, we patterned in the hBN flake an array of spots using the Xe FIB with various fluences. The confocal scan of the patterned hBN region is shown in Figure 2a. The highest and lowest fluence were $1\times10^{17}$ ions/cm² and $2.5\times10^{12}$ ions/cm², respectively. Figure 2b shows the PL intensity vs. Xe fluence measured from the various spots; the inset shows the actual PL spectra recorded for each fluence. The photoluminescence intensity increases as the implantation fluence increases.

The lowest fluence which showed a detectable PL signal from $V_B^-$ defect in our confocal setup (air objective, 0.9 NA, 532-nm laser excitation at 2 mW of power) was $2.5\times10^{13}$ ions/cm². We did not perform annealing of the sample after patterning, for annealing is known to dissociate the $V_B^-$ defects. This effect is known to occur for other vacancy defects, e.g. in silicon carbide.[38]

Next, we studied the formation of $V_B^-$ defects by using atomic species for the ion beam other than xenon, namely argon and nitrogen. We used Stopping and Range of Ions in Matter (SRIM) simulations to calculate the depth and number of vacancies created by the different ion sources at 30 keV. These three species were selected as they are available in commercial plasma-FIB systems and can give further insight into the vacancy formation and related luminescence of the $V_B^-$ defect. The estimated number of vacancies and penetration depth are substantially different for the different ions, as shown in Figure 2c. The analysis indicates that Xe ions are more likely to create vacancies at a shallower depth (peaking at a depth $t \sim 15$ nm) than argon and nitrogen ions (peaking at $t \sim 25$ nm and $t \sim 60$ nm, respectively).

Creation depth is an important parameter for many applications which rely on the optically-active defect being close to the surface, e.g. spin-based sensing applications, for which proximity to the surface improves sensitivity.[6] Intuitively, lighter atoms create vacancies in deeper regions with a significantly broader distribution, while the number of generated vacancies per primary ion is lower. Therefore, a nitrogen FIB can be used to engineer a thicker hBN flake with a homogeneous distribution throughout its volume, while a Xe beam may be utilized to generate defects mostly at the top layers.

The formation of $V_B^-$ defects by the three ions is analyzed employing confocal microscopy. We implanted three separate areas of a hBN flake with the same fluences of Xe, Ar and N, all with an energy of 30 keV. The different ions yielded different intensities for the measured photoluminescence. The $3\times10^{16}$ ions/cm$^{-2}$ N-irradiated area displayed a higher PL intensity compared to the corresponding areas implanted with Xe and Ar (Figure 2d). We attribute this to the nitrogen ions creating a more homogeneous distribution of single vacancies, thus a greater number of $V_B^-$ defects, throughout a relatively larger volume of material (Figure 2c). In fact, the N-implantation seem to induce less of the sputtering and material amorphization we observed for the Xe implantation. While we maximized the luminescence in a given flake, we were unable to observe emission from an isolated defect, even at the lowest fluence of $2.5\times10^{13}$ ions/cm$^2$, from any of the three species.

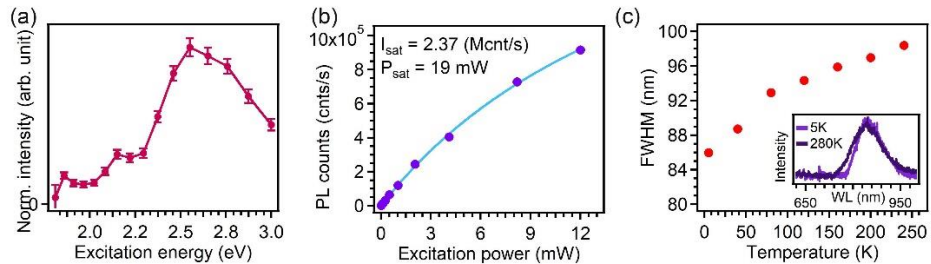

**Figure 3.** Photoluminescence characterization of $V_B^-$ in hBN. **a)** Photoluminescence excitation measurement of an ensemble of $V_B^-$ defects in the range 1.8–3 eV. **b)** PL measurement of $V_B^-$ emission as a function of excitation power. The saturation power of 19 mW is derived from the fit (light blue line). **c)** FWHM of the PL emission as a function of temperature. In our measurement we did not observe any narrow peak within the emission wavelength. The PL spectra at 4 K and 280 K is shown in the inset for reference.

To gain more insight into the photophysical properties of the $V_B^-$ defect, we also performed photoluminescence excitation (PLE) measurements using a tunable laser. The excitation energies were tuned from 3 eV (400 nm) to 1.8 eV (680 nm), while the overall power density on the sample was kept constant. We monitored the PL emission at 820 nm as we scanned through the spectral range of excitation wavelengths. Figure 3a shows the PLE measurement for the $V_B^-$ defects created with the $1\times10^{15}$ ions/cm$^2$ Xe-irradiation. The most efficient excitation occurs at ~2.6 eV (~480 nm), relatively far from the main emission line at 820 nm. Interestingly, at excitation energies below 2 eV (above 620 nm), the excitation of the defect seems inefficient.

Next, we measured the saturation behavior of the $V_B^-$ defects. We excited the emitters using a 532-nm laser while increasing the excitation power to up to 12 mW. The data is shown in Figure 3b. Fitted to the equation: $I = (I_{sat}P)/(P + P_{sat})$, the dataset gives a saturation power of 19 mW corresponding to 2.37 MCounts/s. The relatively large value for $P_{sat}$ suggests that the $V_B^-$ center might possess a low quantum efficiency, consistent with the aforementioned inability to detect photoluminescence from a single, isolated defect. On the other hand, it is important to note that even under 12 mW of excitation power the defects remained stable, with no evident bleaching. Lastly, we performed photoluminescence spectroscopy measurements as a function of temperature to try to identify the energy of the zero-phonon line (ZPL) transition. The emission remains mostly unchanged even at cryogenic temperatures (~5 K). In fact, the linewidth narrows only slightly from ~95 nm at room temperature to ~86 nm at 5 K (Figure 3c). Note that this does not prevent from observing optically detected magnetic resonance. Nevertheless, further research is needed to try to understand the origin of the PL emission for the $V_B^-$ defect.

To conclude, we have demonstrated deterministic engineering (spectrally and spatially) of optically-active $V_B^-$ defects in hBN through ion implantation. The $V_B^-$ defects exhibit room temperature ODMR, and hence are promising candidates to be explored as spin-addressable systems for quantum and sensing applications. We measured the maximum absorption for these defects to be ~2.6 eV (~480 nm) which is ~1 eV higher than the emission maximum. Although the emission gets marginally narrower at cryogenic temperatures, we could not unequivocally identify the zero-phonon line emission energy. Further research is needed to fully understand the photophysics of $V_B^-$ defects in hBN, and isolate single emitters. Our results are an important step towards the controlled engineering of spin defects in hBN. It may prove van der Waals materials as a possible platform—in addition to established ones such as diamond and silicon-carbide—for the realization of spin-based applications in quantum information and quantum sensing. One of the key advantages is the possibility to engineer, on demand, spin defects directly interfacing with prefabricated photonic and phononic hBN crystals to explore spin optomechanics with hBN. Furthermore, the two-dimensional nature of hBN allows for the creation of these spin-defects within a few atomic layers from the surface of the material, which is a critical requirement for applications in quantum sensing.

**Acknowledgements**
The authors thank the Australian Research Council for the financial support (DP180100077, DP190101058, DE180100810).